\documentclass[preprint2]{aastex6} 
\usepackage{CJKutf8}
\newcommand{\myemail}{chunlin.tian@sdu.edu.cn}
\shorttitle{Numerical simulations of KH instability}
\shortauthors{Tian and Chen}
\begin{document}
\title{Numerical simulations of Kelvin-Helmholtz 
    instability: a two-dimensional parametric study}
\author{Chunlin Tian 
        \begin{CJK*}{UTF8}{gbsn} (田春林) \end{CJK*}
         and Yao Chen 
        \begin{CJK*}{UTF8}{gbsn} (陈耀) \end{CJK*}
}
\affil{Shandong Provincial Key Laboratory of Optical Astronomy
           and Solar-Terrestrial Environment,
           Institute of Space Sciences,
           Shandong University, Weihai, 264209, China}
\email{\myemail}
\begin{abstract}
Using two-dimensional simulations, we numerically explore the 
dependences of Kelvin-Helmholtz instability upon various physical 
parameters, including viscosity, width of sheared layer, flow 
speed, and magnetic field strength. In most cases, a multi-vortex 
phase exists between the initial growth phase and final single-vortex 
phase. The parametric study shows that the evolutionary properties, 
such as phase duration and vortex dynamics, are generally sensitive 
to these parameters except in certain regimes. An interesting result 
is that for supersonic flows, the phase durations and saturation of 
velocity growth approach constant values asymptotically as the sonic 
Mach number increases. We confirm that the linear coupling between 
magnetic field and Kelvin-Helmholtz modes is negligible if the magnetic 
field is weak enough. The morphological behaviour suggests that the 
multi-vortex coalescence might be driven by the underlying wave-wave 
interaction. Based on these results, we make a preliminary discussion 
about several events observed in the solar corona. The numerical models
need to be further improved to make a practical diagnostic of the
coronal plasma properties.

\end{abstract}
\section{Introduction}
The velocity shear concentrating in a thin layer is ubiquitous in
natural flows. Under certain circumstance, the velocity shear is 
susceptible to the Kelvin-Helmholtz (KH) instability, and may eventually 
develop into turbulence or large-scale wavy motions. The KH instability 
is a very important mechanism for momentum and energy transport and mixing 
of fluid. The phenomena akin to KH instability are frequently observed 
in the atmosphere of planets, magnetosphere boundary, and low solar 
corona. For instance, ripples at the prominence surface 
(\citealt{ryutova2010}), billows on the flank of a coronal mass 
ejecta (CME) (\citealt{foul2011}), and traveling fluctuations at the 
boundaries of the magnetic structures (\citealt{ofman2011,mostl2013}), 
were attributed to the KH instability. Plasma blobs emerging from the 
cusps of quiescent coronal streamers have also been interpreted as the 
result of nonlinear development of streaming KH instability 
(\citealt{chen2009}).

A comprehensive understanding of the KH instability is useful for 
diagnosing the plasma conditions of the occurring sites. Linear analysis 
gave the onset condition of the KH instability in magnetized plasma 
(e.g., \citealt{chand1961}),
\begin{equation}
[\vec{k}\cdot(\vec{u_1}-\vec{u_2})]^2 >
\frac{\rho_1+\rho_2}{\mu_0\rho_1\rho_2}
[(\vec{k}\cdot\vec{B_1})^2+(\vec{k}\cdot\vec{B_2})^2],
\label{criterion}
\end{equation}
where subscripts indicate the quantities of either side of the velocity 
shear. $\vec{k}$, $\vec{u}$, $\rho$, $\vec{B}$, $\mu_0$ are the wave 
vector, velocity, density, magnetic field, and permeability in vacuum, 
respectively. This criterion is obtained based on the assumption that 
the sheared layer is infinitely thin and the fluid is incompressible. 
\cite{miura1982} showed that compressibility can stabilize KH modes and 
the finite thickness of velocity shear $\Delta$ acts as a filter, e.g., 
only modes with $k\Delta<2$ are unstable and the fastest growing modes 
are those with $k\Delta\sim 0.5-1$. \cite{wu1986} compared the growth 
rate of convective and periodic KH instability and found no difference 
in the linear stage.  

Criterion~(\ref{criterion}) and many other studies indicate that 
the component of magnetic field parallel to the flows can stabilize
KH modes. If magnetic reconnection is taken into account, 
the situation is more complicated. \cite{chenq1997} 
studied the coupling between KH and tearing instability. No linear 
interaction has been confirmed. \cite{nykyri2001} numerically 
investigated the mass transport due to reconnection in KH vortices. 
They found that reconnection would cause the high density plasma 
filaments detached from the magneto-sheath. Similar numerical 
simulations (e.g., \citealt{otto2000} and \citealt{nykyri2006}) 
have been performed to identify and reproduce the processes which 
cause the fluctuations at the boundary between magnetosphere and 
magneto-sheath observed in-situ.

Based on the detailed observational analysis of \cite{foul2013}, 
\cite{nykyri2013} conducted a magnetic seismology study to 
parametrically determine the field configuration in the
CME reconnection outflow layer. Since the magnetic flux rope could 
be a component of CME, the KH instability at a cylindrical surface
is of particular interests. \cite{zhely2015b,zhely2015a} showed
that the KH wave observed in CME is the $m=-3$ MHD mode in the twisted 
flux tube, where $m$ is the azimuthal wave number. Using a
two-fluid approximation, \cite{mart2015} studied the effects of 
partial ionization in cool and dense magnetic flux tubes. 

\cite{frank1996}, \cite{jones1997}, \cite{jeong2000}, and 
\cite{ryu2000} carried out a series of numerical simulations of the 
KH instability. The control parameters they used are magnetic field 
strength, magnetic field orientation, the sheared magnetic fields, 
and the perturbation from third dimension. They found that in the 
hydrodynamic case, the KH vortex persisted until the viscosity 
dissipated it. In the case with magnetic field, they classified 
four parameter regimes: the dissipative, disruptive, nonlinearly 
stable, and linearly stable regime. If the field strength is less than 
$\sim 0.1$ of the critical value needed for linear stabilization, 
the role of magnetic field is to enhance the rate of energy dissipation. 
The magnetic tension force in stronger field cases would disrupt 
the KH vortex. For magnetic field strength slightly weaker than 
required for linear instability, the magnetic tension enhanced during 
linear growth phase prevents the flow from developing into nonlinearly 
unstable state. In the linearly stable regime, only very small 
amplitude fluctuations present. \cite{jones1997} carefully 
designed a set of numerical experiments to show that the magnetic
field component perpendicular to the flow affects the KH instability 
only through minor pressure contributions and the flows are essentially
two-dimensional (2D). Three-dimensional (3D) numerical simulations
indicate that the KH vortex has a quasi-2D structure at the
beginning. For weak magnetic fields, the 3D KH instability will 
eventually develop into decaying turbulence. If the magnetic field is 
relatively strong, the flows will undergo reorganization and become
stable. \cite{mala1996} identified three stages in the evolution of KH 
instability for marginally supersonic weak field case, namely, a 
linear stage, a dissipative transient stage, and a saturation stage.

In most of the earlier simulations, only the dependence of
single-vortex dynamics on magnetic field configuration was 
parametrically studied and the sonic Mach number was fixed. The 
discussions about viscosity, width of velocity shear, and multiple
vortices interaction in these literatures are limited. In the present 
study, besides magnetic field strength, we systematically explore the 
viscosity, width of velocity shear, and flow speed in a wide range 
of values. We are interested in not only the vortex dynamics but also
some direct observable quantities, e.g., the duration of evolution
phases. As expected, we repeat many aspects of the existing studies. 
We also obtain some new results that are discrepant and complementary 
to the earlier investigations.

In the aforementioned KH instability studies, the geometry configuration
of sheared layer is simply an interface. Another type of configuration, 
which could mimic streamer or jets, has also been considered in 
some investigations (e.g., \citealt{min1997,zaliz2003,betta2006,betta2009}). 
Here we only consider the interface geometry in a 2D computational domain.

The rest of this paper is organized as follows. Section \ref{secnum} 
describes the numerical model. Section \ref{secrsl} presents the results 
from numerical simulations. In section \ref{secdisc}, we discuss the 
applications of our results to the observations in solar 
corona. Section \ref{seccon} summarizes and concludes this paper.
\section{Numerical model}
\label{secnum}

We numerically solve the following resistive MHD equations using the 
PENCIL CODE\footnote{https://github.com/pencil-code/}, which
is an open source, modular high-order finite difference code. It is of 
sixth-order accuracy in space and third-order in time by default. 
\begin{eqnarray}
\label{ns1}
\partial\ln{\rho}/\partial t &=& -\vec{u} \cdot \nabla \ln{\rho}
                                 -\nabla \cdot \vec{u},\\
\label{ns2}
\partial\vec{u}/\partial t &=& -\vec{u}\cdot \nabla \vec{u} 
                            -c^2_s\nabla(s/c_p+\ln{\rho})\nonumber\\
                  & & +\nu(\nabla^2\vec{u}+\nabla\nabla\cdot\vec{u}/3
                      +2\vec{\Sigma}\cdot\nabla\ln{\rho})\nonumber\\
                  & & +\zeta\nabla\nabla\cdot\vec{u}
                      +\vec{j}\times\vec{B}/\rho,\\
\label{ns3}
\partial s/\partial t &=&-\vec{u}\cdot\nabla s
           +1/(\rho T)(\nabla\cdot(K\nabla T) \nonumber\\
 && +2\rho\nu\vec{\Sigma}\otimes\vec{\Sigma}
+\zeta\rho(\nabla\cdot\vec{u})^2\nonumber\\
           & & +\eta\mu_0\vec{j}^2),\\
\label{induction}
\partial\vec{A}/\partial t &=& \vec{u}\times\vec{B}
-\eta\mu_0\vec{j},
\end{eqnarray}
where $c_s$ is the sound speed, $c_p$ the specific heat at constant 
pressure, $s$ the specific entropy, $\vec{j}=\nabla\times\vec{B}/\mu_0$ 
the electric current density, $\vec{\Sigma}$ the rate-of-shear tensor 
that is traceless, $\nu$ the kinematic viscosity, $\zeta$ the bulk 
viscosity, $K$ the thermal conductivity, $\vec{A}$ the vector potential, 
and $\eta$ the magnetic resistivity. Other symbols have their 
standard meanings.  

We consider periodic flows in the x-y plane. An ideal gas
is confined in a rectangular computational domain, with $x\in[-2L,2L]$, 
$y\in [0,4L]$. Periodic boundary conditions are used at the y boundaries.
Open boundary conditions are applied to the x boundaries.  All the 
simulations are performed on a $512\times 512$ mesh.

The background plasma is uniform in density $\rho_0$ and 
specific entropy $s_0$. A velocity shear layer with hyperbolic 
tangent profile is initially centered at the y axis ($x=0$), i.e.,
\begin{equation}
u_y=\frac{1}{2}u_0(1+\tanh{\frac{x}{a}}),
\end{equation}
where $a$ is the half width of the sheared layer, and $u_0$ is the 
initial flow speed. A small perturbation is introduced to $u_x$ at
$t=0$ of the form
\begin{equation}
u_x=u'_{x0}(e^{-\frac{(y-y_1)^2}{\Delta_y^2}}
           -e^{-\frac{(y-y_2)^2}{\Delta_y^2}}),
\end{equation}
where $u'_{x0}$ is the amplitude of perturbation, $y_1$, $y_2$,
and $\Delta_y$ define its location and width along the y-direction. 
For magnetized case we impose a uniform magnetic field parallel to 
the initial velocity, i.e., $B_y=B_0$.  

We choose a case with parameters of moderate value as our reference 
model. Then we vary the control parameters to check their influence 
on the dynamics of KH instability. For the reference model, $\rho_0=1$ 
and the magnetic field is absent. The plasma stays still in the left 
half computational domain ($x<0$) and flows with $u_0=0.5$ in the 
right half ($x>0$). The 
initial entropy is set so that $c_s=1.29$, and thus we have 
$Ma=u_0/c_s\simeq 0.39$. With $a=0.05L$, there are nearly 
$7$ grids in the sheared layer.

Dimensionless quantities, namely the thickness of the sheared layer 
$a/L$, sonic Mach number $Ma=u_0/c_s$,  Alfv\'enic Mach number 
$M_A=u_0/c_a$, plasma beta $\beta=p_g/p_m$ , Reynolds number 
$Re=u_0L/\nu$, magnetic Reynolds number $Rm=u_0L/\eta$,  and P\'eclet 
number $Pe=c_p\rho u_0L/K$, are used to characterize the simulation runs.
Considering the stabilization due to magnetic field, only the 
component parallel to the flows takes effect. Therefore,
it is useful to define an effective Alfv\'enic Mach number and an
effective plasma $\beta$ with $M_{A,y}=u_0\sqrt{\mu_0\rho}/B_y$ and 
$\beta_{y}=2\mu_0p_g/B_y^2$.

The reference model is labeled as Run A and listed in Table~\ref{runs} 
among other representative cases. In all runs presented here, 
we adopt $Pe=10^6$ and $Rm=5\times 10^7$. The ranges of other
parameters are listed in the last column of Table~\ref{ranges}.

\begin{table}
 \begin{center}
 \caption{Representative simulation runs. }

 \label{runs}
 \begin{tabular}{@{}lcccccc}
  \hline
  \hline
  Run & $a/L$  &  $Ma$ & $M_{A,y}$ & $Re$   & $\beta_y$  \\
  \hline
  A   & $0.05$ & $0.39$& $--$    &  $2500$ & $--$ \\
  B   & $0.1 $ & $0.39$& $--$    &  $2500$ & $--$ \\
  C   & $0.05$ & $0.39$& $--$    &  $250$  & $--$ \\
  D   & $0.05$ & $0.04$& $--$    &  $2500$ & $--$ \\
  E1  & $0.05$ & $0.39$& $50$    &  $2500$ & $20000$ \\
  E2  & $0.05$ & $0.39$& $3.57$  &  $2500$ & $102$ \\
  E3  & $0.05$ & $0.39$& $2.5 $  &  $2500$ & $50$ \\
  E4  & $0.05$ & $0.39$& $2.27$  &  $2500$ & $41.3$ \\
  \hline
 \end{tabular}
\end{center}
\end{table}

\section{Results}
\label{secrsl}
\begin{figure}
\centering
\includegraphics[scale=1]{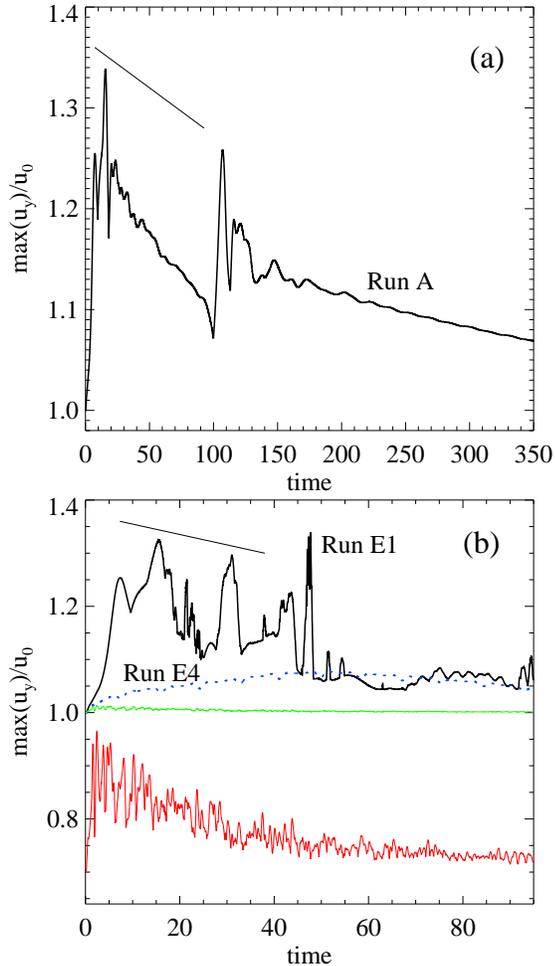}
\caption{
Evolution of $\max{(u_y)/u_0}$ in typical cases. (a) hydrodynamic case
(Run A); (b) thick solid black line, dotted blue line, and thin solid 
green line represent very weak field case (Run E1), the critical case
for linear stability (Run E4), and strong field case, respectively. 
Oblique straight lines indicate the multi-vortex phase in Run A and Run E1.
The bottom red curve in (b) is the green line shifted downward and 
multiplied by a factor of 20 to reveal the small-amplitude fluctuations.
}
\label{fighistory}
\end{figure}

Figure~\ref{fighistory}(a) shows the time evolution of $\max{(u_y)}$ 
in the reference Run A, from which we can see that the evolution 
consists three phases: (i) an initial growth phase, in which
the perturbed flow grows until the vortex starts to form; (ii) a 
multi-vortex stage, in which multiple fully developed KH vortices
coexist until they start to merge; and (iii) a single-vortex stage, in
which the single-vortex resulted from multi-vortex coalescing
spins until the end of the simulation. Some control parameters can
dramatically affect the dynamics of the KH vortex. 

The influence of magnetic field is shown in the panel (b) of 
Fig.~\ref{fighistory}, where the disruption effect of very weak field and 
stabilization effect of strong field are evident. 

In the remainder of this section, besides vortex dynamics, we inspect 
the parametric behaviour of evolutionary phase durations. 
The first peak in the evolutionary curve of $\max{(u_y)}$ marks the 
saturation of the velocity growth. We define the occurrence of this local 
maximum as the end of initial growth phase. The duration of multi-vortex 
phase, $\Delta t_{mult}$, is defined as the time interval between the KH 
vortices evidently form and they start to coalesce. 

\subsection{Initial growth phase}
The initial growth phase can be roughly divided into two stages, 
i.e., a linear stage followed by a nonlinear stage. The KH vortex only 
starts to form in the late nonlinear stage. In this stage the growing 
amplitude of velocity is comparable to the background flow speed, and 
thus the nonlinear effects cannot be ignored anymore. 

\begin{figure*}
\centering
\includegraphics[scale=1]{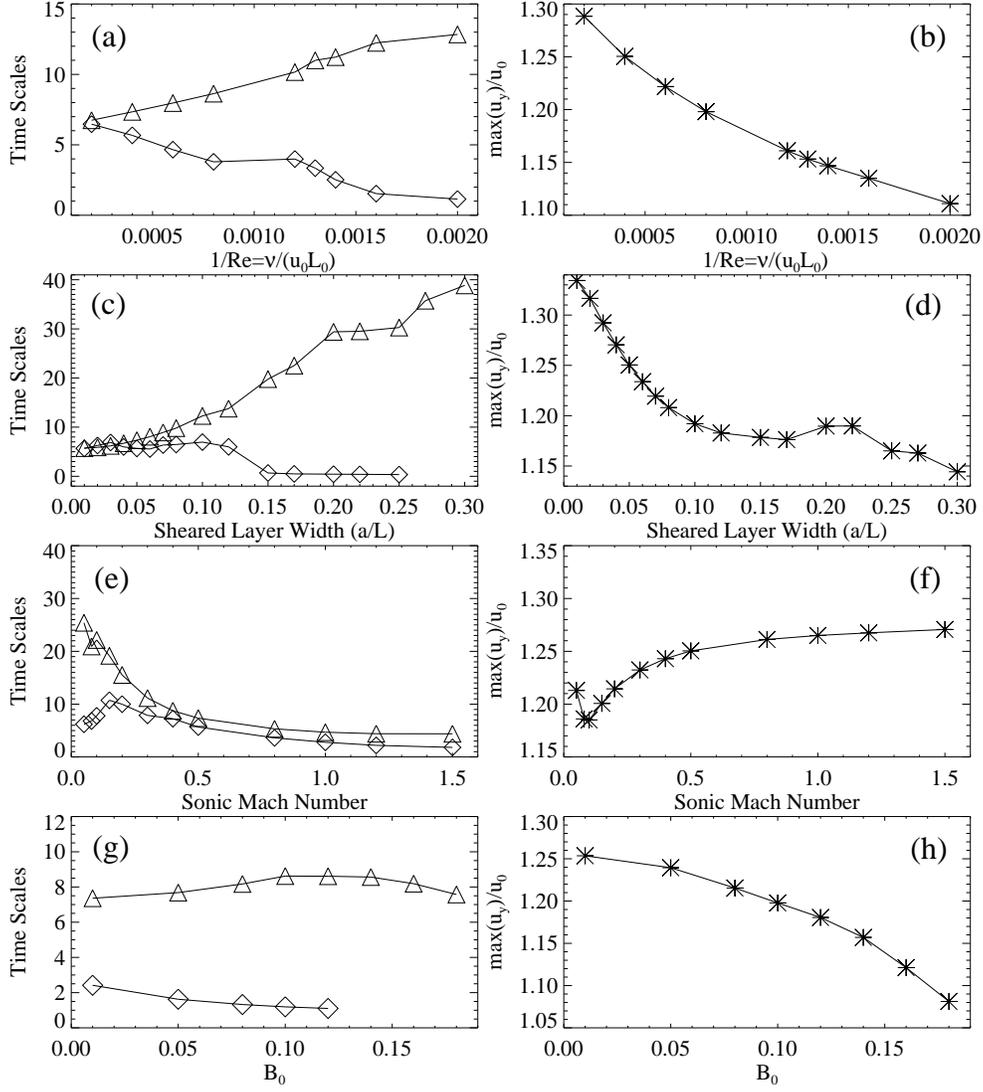}
\caption{
Phase durations and velocity saturation as a function of various
parameters. Stars: velocity saturation which is measured by the first 
peak value of $\max{(u_y)}$; diamonds: the duration of multi-vortex 
phase; triangles: the duration of initial growth phase. Note that the 
period of multi-vortex phase is multiplied by a factor of 1/15.  
In order to linearly space the data points, we plot the curves against 
$1/Re$ instead of $Re$ in panel (a) and (b), and against initial magnetic 
field strength $B_0$ instead of plasma $\beta$ in panel (g) and (h).
}
\label{figrowth}
\end{figure*}

Figure~\ref{figrowth} shows the dependence of initial growth duration, 
$\Delta t_{init}$, on the control parameters. Viscosity and the width of 
sheared layer can affect considerably the initial growth time-scale. It 
is expected that the flow speed is critical to $\Delta t_{init}$. 
But this is true only for the subsonic flows ($Ma\lesssim 0.5$). It is 
interesting to notice that $\Delta t_{init}$ is slightly dependent on the 
speed of supersonic flows and approaches a constant as  sonic 
Mach number increases (see Fig.~\ref{figrowth}(e)).
 As represented by the critical case  in the 
panel (b) of Fig.~\ref{fighistory},  the initial growth phase can be 
hardly identified in the presence of strong magnetic field. 
Figure ~\ref{figrowth}(g) shows that the initial growth duration 
measured in the weak field cases only slightly depend on 
the magnetic field strength. The reason may lie in the fact that the 
KH modes are stabilized by magnetic tension force. During the initial 
growth phase, especially the linear stage, there is no obvious vortex 
formed, and thus no Maxwell stress is induced by magnetic field 
distortion. So the strength of weak magnetic field plays a minor role 
in determining $\Delta t_{init}$. It should be pointed out that 
here we only talk about the growth of velocity. By contrast, the growth of 
magnetic field is sensitively dependent on the initial strength. 
 
Both $u_x$ and $u_y$ are enhanced continuously by Reynolds stress
during the initial growth phase.  Panels on the right of 
Fig.~\ref{figrowth} show the value of the first peak of $\max(u_y)$  in
its evolution curve as a function of various parameters.  
For the present study, in many cases the saturation of velocity 
enhancement is between $15\%$ and $30\%$
of the background flow speed. The first peak decreases monotonically 
as a function of viscosity and magnetic field strength. 
We identify a parameter range,  $0.1\lesssim a/L \lesssim 0.2$,  
in which the first peak varies slightly.  Compare the panel (e) and (f)
in Fig.~\ref{figrowth}, we can see the similarity between $\Delta t_{init}$
and  $\max(u_y)/u_0$.  The first peak of $\max(u_y)/u_0$  
is nearly independent on $Ma$ for supersonic flow and approaches 
an asymptotic upper limit of $\sim 1.27$ as the sonic Mach number
increases. 

\subsection{Multi-vortex phase}
\begin{figure}
\centering
\includegraphics[scale=1]{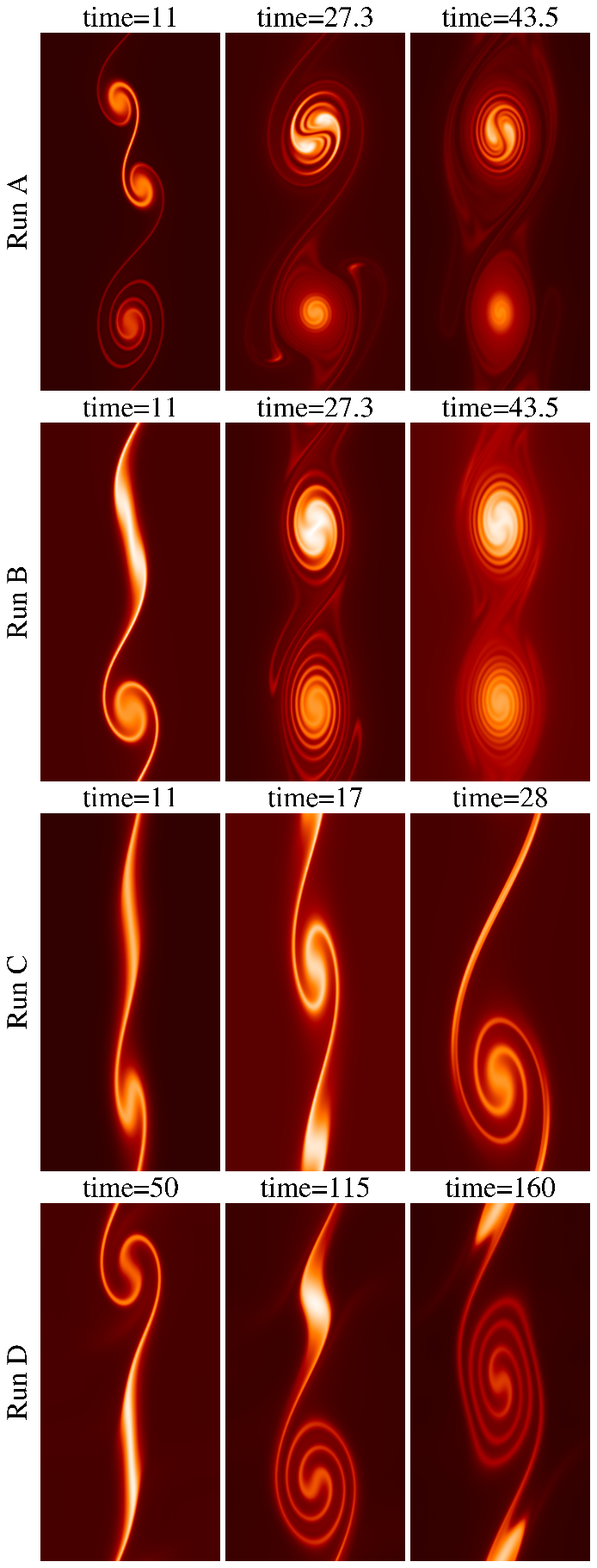}
\caption{
Snapshots of specific entropy taken during multi-vortex evolving phase 
for selected cases, i.e., Run A (the reference case), Run B (with wider 
sheared layer width), Run C (with larger viscosity) and Run D (with 
smaller sonic Mach number). 
}
\label{figmulti1}
\end{figure}

\begin{figure}
\centering
\includegraphics[scale=1]{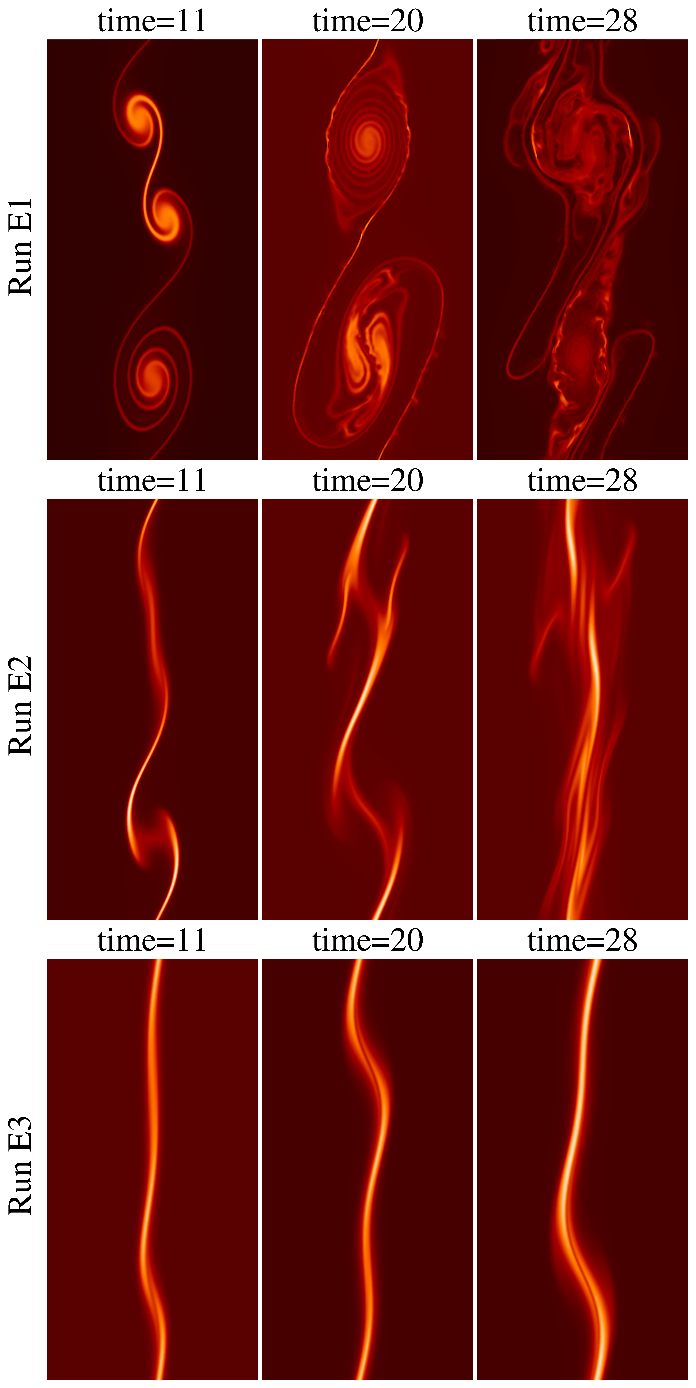}
\caption{
Snapshots of specific entropy taken during multi-vortex evolving phase 
for selected cases with different effective Alfv\'enic Mach number. Run E1: 
$M_{A,y}=50$; E2: $M_{A,y}=3.75$; E3: $M_{A,y}=2.5$. 
}
\label{figmulti2}
\end{figure}

The number of KH vortex is determined by the initial perturbation.
In our models, two Gaussian perturbations with half width of $0.1L$ 
are introduced. After the initial growth, the KH vortex with 
$\lambda=2L$ form and evolve to $\lambda=4L$ later on. The
multi-vortex phase can also be defined as the evolutionary phase 
with $\lambda=2L$. Figure~\ref{figmulti1} and Figure~\ref{figmulti2} 
show the snapshots of selected typical cases during the multi-vortex 
phase. The parameters used in these runs are given in Table~\ref{runs}.
 
In the present simulations, there are two fully developed vortices in 
most of the runs. The coexisting vortices are generally different 
in appearance. For example, in the reference Run A, the relatively 
round vortex resembles the `yin-yang symbol' in Chinese traditional 
philosophy and the more oval vortex looks just like Cat's Eye. 
Inspection reveals that the `yin-yang symbol' is resulted 
from pairing process. We will discuss this issue in the next 
subsection. If the sheared layer is moderately wide ($a/L=0.1$), these 
two vortices are similar to each other (see Run B). In a thinner 
sheared layer model ($a/L=0.01$), both vortices have the appearance of 
`yin-yang symbol'. Under certain conditions,  for instance,
the flow is very viscous (see Run C), very slow (see Run D) or the velocity
shear is very wide ($a/L\gtrsim 0.2$), the `yin-yang symbol' vortex cannot 
develop. 


The multiple vortices spin until they merge into a single vortex.   Figure~\ref{figrowth} also shows the 
estimated durations of multi-vortex phase. Note that in some cases, 
the multi-vortex phase cannot be clearly identified. The time-scale of 
multi-vortex phase is very sensitive to the parameter ranges
explored in the current study, except for $Re\sim 1000$ and 
$a/L\sim 0.08$. Usually the multi-vortex phase lasts longer than the
initial growth phase but an exception exists for $a/L\gtrsim 0.15$. 
For supersonic flows, the multi-vortex phase duration
$\Delta t_{mult}$ approaches asymptotically a constant just like
$\Delta t_{init}$.

The presence of magnetic field alters dramatically the KH vortex
patterns, and thus the multi-vortex phase is recognizable  only
if the magnetic field is sufficiently weak (see Fig.~\ref{figrowth}(g)). 
Compared to the 
hydrodynamic case, the presence of magnetic field can speed up 
the multi-vortex phase by a  factor of 2 in the very weak field case
($\beta_y \sim 20000$).  The multi-vortex phase duration 
decreases as the magnetic field strength increases. 

\subsection{Multi-vortex coalescing}
\begin{figure*}
\centering
\includegraphics[scale=1]{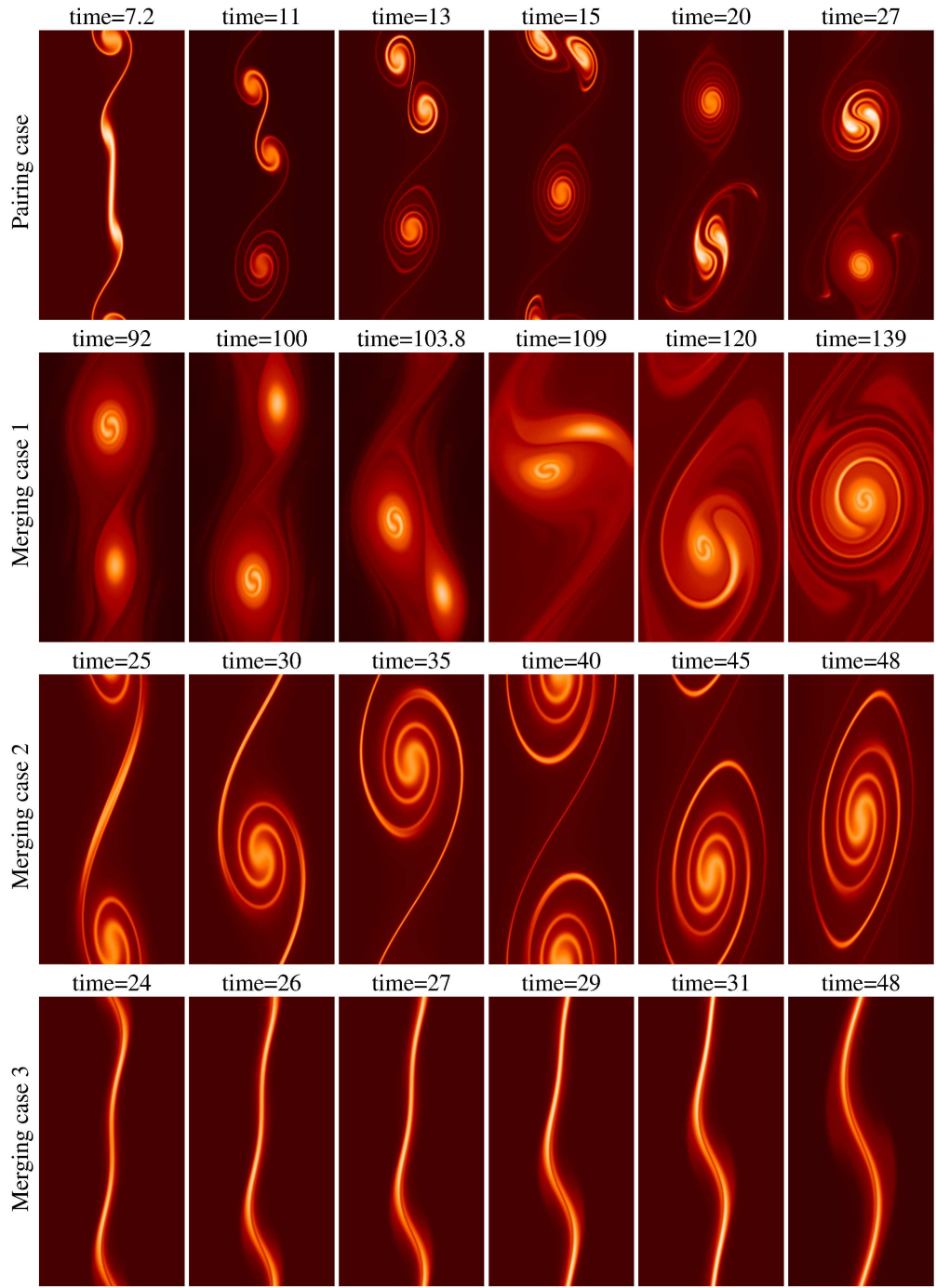}
\caption{
Coalescing processes taking place at different stages in the typical cases.
Top panels: pairing process at the beginning of multi-vortex phase
in Run A; Merging case 1: wrapping process at the end of multi-vortex 
phase in Run A;  Merging case 2: merging process in Run C, where 
the `yin-yang symbol' vortex cannot develop; Bottom panels: merging of 
the wavy motion in Run E3. 
}
\label{figcoalesce}
\end{figure*}

It has been pointed out that coexisting KH vortices would merge 
(e.g., \citealt{frank1996}) and the vortex pairing process transfers
energy from short wavelength to long wavelength perturbations
(e.g., \citealt{mala1996}). The present numerical simulations show 
that the KH vortices merge through pairing or wrapping process. 
Figure~\ref{figcoalesce} shows the typical coalescing processes 
for selected cases.

Pairing takes 
place at the beginning stage of multi-vortex evolution phase. 
Top panels of Fig.~\ref{figcoalesce} show that two newborn vortices 
rotate symmetrically around each other during pairing process.
Eventually, a bigger vortex forms in the shape 
of `yin-yang symbol'. The two engaged vortices are symmetric through
the period of pairing. 

The merging case 1 in  Fig.~\ref{figcoalesce}  represents a typical 
wrapping process. Wrapping happens near the end of the multi-vortex 
evolution.  During the wrapping process, the Cat's Eye eddy shrinks 
continuously and then is wrapped to the `yin-yang 
symbol' eddy. The resulted single-vortex is bigger and more 
complicated in the fine structure. The Cat's Eye eddy becomes a 
part of the perimeter structure.  The merging case 2 in Fig.~\ref{figcoalesce} 
is a special wrapping process occurring in the very viscous flows.
In this case, the `yin-yang symbol' vortex cannot develop and
the shearing layer outside the Cat's Eye vortex is wide at first. 
During the course of merging, the outside shearing layer becomes 
thinner and is finally wrapped to the Cat's Eye vortex rapidly. 
Afterwards, the size of the Cat's Eye is doubled and comparable 
to the computational domain.  

As long as the vortex formation is suppressed by the magnetic 
tension force, the KH instability develops into wavy motions.
The merging case 3 in Fig.~\ref{figcoalesce} can be regarded
as a process of  `multi-vortex coalescing' in this special circumstance.
During the coalescing course, the wave-like motion with wavelength 
$\lambda=2L$
evolves into larger structure with $\lambda=4L$ (see Run E3).
Comparing these different merging cases,
we may postulate that the multi-vortex coalescing is driven 
by the underlying wave-wave interaction. When the vortex is the 
dominant feature of the KH instability, the coalescing process is
manifested by the complicated vortex dynamics.

\subsection{Role of uniform magnetic field}
The effects of magnetic field on the KH instability have been studied
intensively with numerical simulations and theoretical analyses. Most of 
the numerical simulations were conducted for weak or very weak fields. 
The present study reproduces many aspects of these results 
(see Fig.~\ref{fighistory}(b)). Here we present some discrepant and 
supplementary results.

We numerically determine the onset condition for the MHD KH  
instability. The critical Alfv\'enic Mach number is $M_{A,y}\sim 2.27$, 
which is a little bit larger than the theoretical value $2$. 
The theoretical prediction is based on several assumptions; for instance, 
the fluid is imcompressible and the sheared layer is infinitely thin. 
Since the compressbility and finite width stabilize the KH modes, 
a little bit larger numerical value is expected. The Alfv\'enic Mach number
can be expressed in term of sonic Mach number and plasma $\beta$, i.e., 
$M_{A,y}\propto Ma\sqrt{\beta_y}$. We conduct a numerical experiment by
varying simultaneously $Ma$ and $\sqrt{\beta_y}$, and keeping their 
product unchanged. The results confirm 
that the linear stability is indeed determined by Alfv\'enic Mach 
number instead of plasma $\beta$. This supports the argument that 
the competition between Maxwell stress and Reynolds stress dominates 
the fate of the KH vortices.

The very weak field dissipative case ($M_{A,y}=50$, $\beta_y=20000$) 
in the present study is different from previous study, for example, 
the simulations done by \cite{jones1997} and \cite{jeong2000}. In 
their study ($\beta=3000$ and $\beta=24000$), the KH  vortex persists 
until viscosity and small-scale magnetic reconnection dissipate it. 
In our case, the KH vortex starts being destroyed soon after the 
small-scale magnetic reconnection taking place, as indicated by the 
top panels in Fig.~\ref{figmulti2}. A possible explanation is that 
the numerical dissipation is significantly different in different 
MHD code. A large dissipation smooths small-scale structures quickly, 
and thus stops the small-scale magnetic reconnection before it 
dramatically disrupts the vortex. We check this effect by adopting 
a very large kinematic viscosity $\nu$ in one of the simulations. 
When the local small-scale velocity gradient is greatly reduced by 
viscosity, the magnetic field cannot be efficiently amplified and thus 
can hardly influence the flow motions. The resulted 
vortex dynamics nearly resembles the hydrodynamic case.

We identify a filamentary structure regime for
$2.8\lesssim M_{A,y}\lesssim 6.2$ ($61\lesssim\beta_y\lesssim 313$). 
An example (Run E2) is shown in Fig.~\ref{figmulti2}. In this regime, 
the magnetic tension force suppresses the swirling vortex when it is
halfway through its first rotation. The half rolled-up vortex extends 
along the field lines to release its kinetic energy and the flow 
pattern becomes filamentary. 

For the stronger field cases ($2.27\lesssim M_{A,y}\lesssim 2.8$), 
the KH modes cannot overcome the magnetic tension force at very 
beginning. The KH instability eventually develops into wavy motions 
(see Run E3 in Fig.~\ref{figmulti2}). The wavelength of these wavy 
structures is finally comparable to the $y$ extent of the 
computational domain. 

\section{Discussion}
\label{secdisc}

\begin{table*}
 \begin{center}
 \caption{Comparison of observational and numerical non-dimensional 
parameters.}

 \label{ranges}
 \begin{tabular}{@{}lccccccc}
  \hline
  \hline
  Parameter& F2011  &  O2011 & M2013 & F2013 & Numerical   \\
  \hline
   $Ma$   & $0.08\sim 1.75$ & $0.12\sim 0.17$ & $0.66\sim 5.66$ & $0.9\sim 3$ & $0.05-1.5$ \\
   $M_A$  & $0.04\sim 2.25$ & $\sim 0.05$ & $0.16\sim 0.68$ & $1.5\sim 5$ & $--$ \\
   $\beta$& $0.16\sim 2.51$ & $\sim 0.15 $ & $\sim 0.22$ & $1.5\sim 9$ & $--$ \\
   $M_{A,y}$& $2.41\sim 12.98$ & $\sim 5$ & $0.8\sim 6.8$ & $--$ & $0.56-50$ \\
   $\beta_y$& $16\sim 854$ & $\sim 1240$ & $5.48\sim 21.95$ & $--$ &  $31-20000$ \\
   $\Delta/\lambda$& $0.17\sim 0.27$ & $\sim 0.11 $ & $0.13\sim 0.2$ & $--$ & $0.01-0.3$ \\
   $Re$  & $--$ & $--$ & $--$ & $--$ & $250-2500$ \\
  \hline
 \end{tabular}
\end{center}
\end{table*}

Our results are 
applicable to a wide variety of astrophysical problems. In this
paper, we present a preliminary application to the solar corona.

\subsection{Applicability to solar corona}
According to the linear analysis, the KH instability may be 
excited by superalfv\'enic flows anywhere in the solar atmosphere 
(\citealt{ryuto2015}). This kind of instability is actually the 
oscillation of flux tube. As a gas dynamics dominated structure, 
the classic rolled-up KH vortex can develop only if plasma $\beta$ 
or sonic Mach number is extremely large. Since the solar corona is 
highly structured and very dynamic (see Fig.~1.17 
in \citealt{aschwan2005}), the strength of magnetic field may vary 
considerably. If we consider the hydrostatic equilibrium, plasma 
$\beta$ varies much faster than magnetic field strength. The 
well-known coronal condition, $\beta\ll 1$, should be applied to the 
magnetic field dominated regions. Far from the major area of these 
regions, e.g., at the interface between different structures, it is 
possible that plasma $\beta$ is considerably larger than unit. It is 
probable that rolled-up KH vortex can develop at these locations 
during some fast transient processes. 

In order to validate the application of our results to the solar 
corona, in Table~\ref{ranges}, we compare the non-dimensional 
parameters used in the current study to that taken or roughly deduced 
from observations. F2011, O2011, M2013, and F2013 stand for the events 
that reported by \cite{foul2011}, \cite{ofman2011}, \cite{mostl2013}, 
and \cite{feng2013}, respectively. Note that the coronal magnetic 
field cannot be directly measured. So the field strength and orientation 
presented in these observations are indeed given by rough estimates. 
Table~\ref{ranges} shows that the plasma conditions for the coronal 
KH instability vary dramatically from case to case, and the parameter
ranges in our study at least partially overlap with the observations. 
Especially the effective Alfv\'enic Mach number lies exactly in the 
theoretically predicted range. 

In a realistic situation, the development of KH instability is 
essentially a 3D problem.  The numerical experiments conducted by 
\cite{ryu2000} indicate that in the early stage of 3D KH instability
development, 2D Cat's Eye develops and is subsequently destroyed 
in all the nonlinearly unstable cases. The fully developed 3D KH
instability is either decaying turbulence for weak field or become
stable for strong field. The 2D rolled-up vortex is the most 
distinguishable feature of the KH instability and thus easily 
identifiable during the observation. So a 2D investigation is of 
practical meaning. 

\subsection{Evolutionary phase duration}

The observed growing and evolving durations of the KH instability 
are significantly different. \cite{foul2013} estimated a period of 
around 2 minute between the first acceleration jet and the first 
perturbation appeared on the CME flank, and the evolution of visible 
vortices lasts about 45 seconds. \cite{ofman2011} obtained a
developing period of 13 minutes, and an evolving period of more
than one and half hours. The event analyzed by \cite{mostl2013}
has estimated growing period of 6 minutes. 

The linear growth rate of the KH mode for a plasma with uniform 
density is  $q=\frac{1}{2}k\Delta v=\pi\Delta v/\lambda$. So 
provided velocity shear and wavelength, we can calculate the 
linear growth rate for the observed events.  Table~\ref{events} 
contains the calculated linear growth rate and some observed 
properties of the coronal KH instability, where $\Delta t_{init}$ 
and $\Delta t_{evol}$ are the growing and evolving duration, 
respectively. Unless the whole growth phase is linear and the 
saturation of velocity is a universal constant, we cannot expect 
that the time-scale of initial growth is uniquely determined by 
the linear growth rate. In order to explain the observations 
we need to consider the nonlinear effect, and the results in 
Fig.~\ref{figrowth} may shed light on it.

In numerical simulations we often have to adopt a Reynolds number 
much smaller than the realistic value by several orders of magnitude. 
As previously stated, the viscosity can affect the initial growth 
period considerably. A linear extrapolation from 
Fig.~\ref{figrowth}(a) suggests that 
$6.5\lesssim\Delta t_{init}\lesssim 6.74)$ when $Re\gtrsim 10000$. 
This means that if the Reynolds number is significantly large, 
the difference in initial growth duration caused by viscosity 
is extremely limited. Also, the magnetic field in the KH events
observed in the low solar corona need to be weak enough so that the 
vortices can roll up.  As mentioned in the Results section, the 
initial growth duration is only slightly dependent on the weak 
magnetic field strength.
These two results suggest that among the tested parameters we 
should concentrate on the width of sheared layer and flow speed 
to explain the differences in the observed coronal KH events.

Firstly we compare the two events observed by \cite{foul2011}
and \cite{mostl2013}. The fast flows in these two events are
supersonic (see Table~\ref{ranges}). According to the results
in Fig.~\ref{figrowth}(e), the initial growth period approaches
an asymptotic constant for supersonic flows. By contrast, we have
$\Delta t_{init}(M2013)/\Delta t_{init}(F2011)=3$ from observations.
The cause of the difference may be the width of the sheared layer.
But the uncertainties in measurement prevent a deterministic  
comparison of the shear width in these two events. 
Using Fig.~\ref{figrowth}(c)  we may roughly estimate that the initial growth 
period varies by a factor of 3 for $0.1\lesssim\Delta/\lambda\lesssim 0.3$.
Since $\Delta t_{init}(M2013)>\Delta t_{init}(F2011)$, we expect 
that the width of velocity shear $\Delta/\lambda$ is wider in M2013.

Then we discuss the difference between F2011 and O2011. 
Table~\ref{ranges} shows that the flow speed in these two 
events is very different. For fast flows, $Ma(F2011)\sim 1.5$ and
$Ma(O2011)\sim 0.15$. From numerical simulations 
(see Fig.~\ref{figrowth}(c)), we roughly have 
$\Delta t_{init} (Ma=0.15)/\Delta t_{init} (Ma=1.5)\sim 4.4$. 
From observations (see Table~\ref{ranges}), we get
$\Delta t_{init} (O2011)/\Delta t_{init} (F2011)\sim 6.5$.
The observed ratio is too large compared to the numerical value. 
This should be from the width of sheared layer. 
Since a wider width causes a longer initial
growth duration, we expect that $\Delta/\lambda$
is larger in O2011. 

For several reasons, we cannot make a similar discussion for the 
multi-vortex evolutionary phase at the present stage. Firstly in
the low solar corona the rapid evolution of background structure 
may eliminate the existing conditions, and thus terminates the KH 
mode before it develops into multi-vortex phase. Secondly with 
nowadays instruments, the fine structure of KH vortices cannot be 
resolved in the low solar corona and we cannot tell if a rolled-up 
vortex has gone through coalescing or not. Another limitation is 
from the numerical models. In the present study, there are only two 
fully developed vortices during the multi-vortex phase. In realistic 
situations, the train of KH vortices may undergo hierarchical merging 
process if the plasma conditions at the occurring place are stable. 
Nevertheless, a diagnostic is possible for some special cases. For 
example, in nonlinearly stable regime, the dynamic vortices are 
suppressed by magnetic field, the multi-vortex evolution can be 
traced by wave-wave interaction. This situation resembles somewhat 
the event observed by \cite{feng2013}. But their observation is made 
in the high corona, and the KH instability triggering event cannot be 
traced. There is no obvious wave-wave interaction in this event either.  
We will discuss this event further in the next subsection.

\begin{table*}
 \begin{center}
 \caption{Properties of observed KH instability in solar corona. }
 \label{events}
 \begin{tabular}{@{}lcccccc}
  \hline
  \hline
  Parameter& F2011  &  O2011 & M2013 & F2013   \\
  \hline
   $\Delta t_{init}$ & $2m$  & $13m$ & $6m$ & $--$ \\
   $\Delta t_{evol}$ & $45$s & $>90m$ & $--$ & $--$ \\
   $\lambda$ & $\sim 18.5\pm 0.5$Mm & $\sim 7$Mm & $\sim 14.4$Mm & $2\sim 3R_{\odot}$ \\
   $h$ & $\sim 10$Mm & $--$ & $2.5\sim 4$Mm & $0.3\sim 0.5R_{\odot}$ \\
   $h/\lambda$ & $0.53\sim 0.56$ & $--$ & $0.17\sim 0.28$ & $0.1\sim 0.25$ \\
   $\Delta v$& $\sim 680$km/s & $6\sim 20$km/s & $\sim 320\pm 40$km/s & $\sim 350$km/s \\
   $q$ & $0.113\sim 0.169$/s & $0.003\sim 0.009$/s & $0.061\sim 0.079$/s & $0.0005\sim 0.00075$/s \\
  \hline
 \end{tabular}
\end{center}
\end{table*}

\subsection{Vortex size}
The height of the billow structure observed in \cite{foul2011} reaches 
$h\sim 10 Mm \sim 0.5 \lambda$. The vortex features observed by 
\cite{ofman2011} are $\sim 7 Mm$ in size, which is also the wavelength
they assumed for analysis. The size of vortices from \cite{mostl2013}
ranges from $\sim 0.17 \lambda$ to $\sim 0.28 \lambda$. The amplitude
of the wave-like motion observed by \cite{feng2013} is $\sim 0.1 \lambda$ 
and increases over the observing period. The current simulations show 
that for mature vortex the ratio of $h/\lambda$ varies approximately 
between $0.25$ and $0.5$. The discrepancy between the numerical
and observed ranges (see Table~\ref{events}) could be caused by the
evolutionary phase difference.  

The KH vortex usually has a oval shape. In some cases, it becomes 
relatively round right after merging, and is elongated along the 
velocity shear lately.  In the event reported by \cite{foul2011},
some vortices rotate about $180$ degree for $\sim 24s$ and then 
disappear probably due to the change of background structures. The 
variation of height of the KH vortices reported by \cite{mostl2013} 
can be attributed to the rotation of the elliptical vortex. The 
identification of rotating vortices suggests that $M_{A,y}\gtrsim 6.2$ 
in these events, otherwise the KH modes will develop into filamentary 
or wavy flow motions.

The event reported by \cite{feng2013} resembles the numerical 
run E3 very closely. In E3, $Ma=0.39$, $\beta_y=50$, 
and $M_{A,y}=2.5$ (see Table~\ref{runs}). In F2013, $Ma\sim 3$
(see Table~\ref{ranges}). Recall that the key parameter for 
KH instability is the effective Alfv\'enic Mach number and 
$M_{A,y}\propto Ma\sqrt{\beta_y}$. Assuming the effective Alfv\'enic
Mach number is roughly same in E3 and F2013, we can estimate
that $\beta_y\sim 50(0.39/3)^2\sim 0.85$ in F2013. From observation
and the general properties of the solar corona, we estimate that
$1.5<\beta<9$ in F2013 (see Table~\ref{ranges}). Considering the 
uncertainties in these estimations, the difference is not that large.

In run E3 the length scale of perturbation is initially $0.2L$
and increases during the development of KH instability. It reaches 
$\sim 2L$ in the multi-vortex phase and $\sim 4L$ after coalescing. 
Meanwhile, the amplitude of the wavy motion increases until it 
reaches $\sim 0.5L$.  In F2013 the length scale and amplitude also 
increase continuously during the observation. Table~\ref{events} shows 
that the wavelength increases from $2R_{\sun}$ to $3R_{\sun}$ and 
the amplitude increases from $0.3R_{\sun}$ to $0.5R_{\sun}$. It seems
that the variation ranges in simulation and observation are close to
each other. But this is not a fair comparison due to the limitations 
of numerical models. In simulations the background plasma is uniform 
and the wavelength is restricted by the $y$ extent of the computational 
domain. In the solar corona the KH wavelength can increases freely and 
the rapid dropping of the background plasma density may cause a fast
growth of the amplitude.

\section{Summary and conclusion}
\label{seccon}

Based on 2D numerical simulations, the dependences of KH instability
on some important parameters have been investigated. The  
parameters that we explored are viscosity, sheared layer 
width, flow speed, and magnetic field strength. In
the present study, we focus on the evolutionary phase duration 
and KH vortex morphology. The main results can be summarized as follows.

For typical hydrodynamic cases, we discern three stages in 
the evolution of the KH instability, i.e., a multi-vortex phase 
which is preceded by a monotonically growing phase and followed 
by a single-vortex spinning phase. The presence of magnetic field 
and variation of parameters may greatly affect these stages in a 
complicated way. For example, the initial growth time scale is 
sensitive to the tested parameters, but there are some regimes, 
such as $a/L\sim 0.2$, $Ma\gtrsim 1$, and  $M_{A,y}\gtrsim 3$,
in which the initial growth duration varies slightly. An interesting 
point from the present simulations is that for supersonic flows,  
the phase durations and saturation of flow growth asymptotically 
approach constant values as the sonic Mach number increases. Although 
magnetic field can dramatically alter the KH vortex morphology, the 
linear coupling between magnetic field and KH modes during the 
initial growth phase is negligible. 

In many cases a KH vortex with appearance of `yin-yang symbol'
is formed through the pairing process. In the pairing process, two
newborn vortices rotate around each other symmetrically and finally
merge into one vortex.  At the end of multi-vortex phase, the KH
vortices coalesce through wrapping process, in which
 one vortex is wrapped to the other and becomes a part of
 the perimeter structure of the resulted single-vortex.
 When the formation of KH vortex is suppressed, the 
coalescence happens between wavy structures;
therefore we may speculate that the multi-vortex coalescing is  
driven by underlying wave-wave interaction and manifested by vortex 
dynamics. 

In our simulations, the MHD KH mode is linearly stable for 
$M_{A,y}\lesssim 2.27$. In the regime $2.27\lesssim M_{A,y}\lesssim 2.8$, 
the MHD KH mode is nonlinearly stable and develops into wavy 
motions. A weak magnetic field with $2.8\lesssim M_{A,y}\lesssim 6.2$ 
causes the KH mode evolving into filamentary flows. The KH vortex
can roll up in an even weaker magnetic field, e.g., a case with
$M_{A,y}=50$. But the small-scale reconnection can destroy the 
integrity of vortex soon after its formation. 

As a fundamental mechanism responsible for various astrophysical 
phenomena, KH instability is of general interests. Based on
the results from 2D numerical simulations, we make a general 
discussion about four events observed in solar corona. It is
promising to develop a practical diagnostic tool for the coronal 
plasma properties. In order to do so, the current numerical KH models
need to be further improved. The plasma $\beta$, sonic Mach number, 
width of sheared layer, and magnetic topology need to be set 
simultaneously according to the properties of the targeted 
phenomenon.

\acknowledgments
This work is funded by China Postdoctoral Science Foundation.
It is also financially supported by NNSFC grants 41274175,
41331068, and NSBRSF grant 2012CB825601. 

\bibliographystyle{aasjournal}
\bibliography{khi}

\begin{thebibliography}{}
\expandafter\ifx\csname natexlab\endcsname\relax\def\natexlab#1{#1}\fi

\bibitem[{{Aschwanden}(2005)}]{aschwan2005}
{Aschwanden}, M.~J. 2005, {Physics of the Solar Corona. An Introduction with
  Problems and Solutions (2nd edition)}

\bibitem[{{Bettarini} {et~al.}(2006){Bettarini}, {Landi}, {Rappazzo}, {Velli},
  \& {Opher}}]{betta2006}
{Bettarini}, L., {Landi}, S., {Rappazzo}, F.~A., {Velli}, M., \& {Opher}, M.
  2006, \aap, 452, 321

\bibitem[{{Bettarini} {et~al.}(2009){Bettarini}, {Landi}, {Velli}, \&
  {Londrillo}}]{betta2009}
{Bettarini}, L., {Landi}, S., {Velli}, M., \& {Londrillo}, P. 2009, Physics of
  Plasmas, 16, 062302

\bibitem[{{Chandrasekhar}(1961)}]{chand1961}
{Chandrasekhar}, S. 1961, {Hydrodynamic and hydromagnetic stability}

\bibitem[{{Chen} {et~al.}(1997){Chen}, {Otto}, \& {Lee}}]{chenq1997}
{Chen}, Q., {Otto}, A., \& {Lee}, L.~C. 1997, \jgr, 102, 151

\bibitem[{{Chen} {et~al.}(2009){Chen}, {Li}, {Song}, {Shi}, {Feng}, \&
  {Xia}}]{chen2009}
{Chen}, Y., {Li}, X., {Song}, H.~Q., {et~al.} 2009, \apj, 691, 1936

\bibitem[{{Feng} {et~al.}(2013){Feng}, {Inhester}, \& {Gan}}]{feng2013}
{Feng}, L., {Inhester}, B., \& {Gan}, W.~Q. 2013, \apj, 774, 141

\bibitem[{{Foullon} {et~al.}(2011){Foullon}, {Verwichte}, {Nakariakov},
  {Nykyri}, \& {Farrugia}}]{foul2011}
{Foullon}, C., {Verwichte}, E., {Nakariakov}, V.~M., {Nykyri}, K., \&
  {Farrugia}, C.~J. 2011, \apjl, 729, L8

\bibitem[{{Foullon} {et~al.}(2013){Foullon}, {Verwichte}, {Nykyri},
  {Aschwanden}, \& {Hannah}}]{foul2013}
{Foullon}, C., {Verwichte}, E., {Nykyri}, K., {Aschwanden}, M.~J., \& {Hannah},
  I.~G. 2013, \apj, 767, 170

\bibitem[{{Frank} {et~al.}(1996){Frank}, {Jones}, {Ryu}, \&
  {Gaalaas}}]{frank1996}
{Frank}, A., {Jones}, T.~W., {Ryu}, D., \& {Gaalaas}, J.~B. 1996, \apj, 460,
  777

\bibitem[{{Jeong} {et~al.}(2000){Jeong}, {Ryu}, {Jones}, \&
  {Frank}}]{jeong2000}
{Jeong}, H., {Ryu}, D., {Jones}, T.~W., \& {Frank}, A. 2000, \apj, 529, 536

\bibitem[{{Jones} {et~al.}(1997){Jones}, {Gaalaas}, {Ryu}, \&
  {Frank}}]{jones1997}
{Jones}, T.~W., {Gaalaas}, J.~B., {Ryu}, D., \& {Frank}, A. 1997, \apj, 482,
  230

\bibitem[{{Malagoli} {et~al.}(1996){Malagoli}, {Bodo}, \& {Rosner}}]{mala1996}
{Malagoli}, A., {Bodo}, G., \& {Rosner}, R. 1996, \apj, 456, 708

\bibitem[{{Mart{\'{\i}}nez-G{\'o}mez}
  {et~al.}(2015){Mart{\'{\i}}nez-G{\'o}mez}, {Soler}, \& {Terradas}}]{mart2015}
{Mart{\'{\i}}nez-G{\'o}mez}, D., {Soler}, R., \& {Terradas}, J. 2015, \aap,
  578, A104

\bibitem[{{Min}(1997)}]{min1997}
{Min}, K.~W. 1997, \apj, 482, 733

\bibitem[{{Miura} \& {Pritchett}(1982)}]{miura1982}
{Miura}, A., \& {Pritchett}, P.~L. 1982, \jgr, 87, 7431

\bibitem[{{M{\"o}stl} {et~al.}(2013){M{\"o}stl}, {Temmer}, \&
  {Veronig}}]{mostl2013}
{M{\"o}stl}, U.~V., {Temmer}, M., \& {Veronig}, A.~M. 2013, \apjl, 766, L12

\bibitem[{{Nykyri} \& {Foullon}(2013)}]{nykyri2013}
{Nykyri}, K., \& {Foullon}, C. 2013, \grl, 40, 4154

\bibitem[{{Nykyri} \& {Otto}(2001)}]{nykyri2001}
{Nykyri}, K., \& {Otto}, A. 2001, \grl, 28, 3565

\bibitem[{{Nykyri} {et~al.}(2006){Nykyri}, {Otto}, {Lavraud}, {Mouikis},
  {Kistler}, {Balogh}, \& {R{\`e}me}}]{nykyri2006}
{Nykyri}, K., {Otto}, A., {Lavraud}, B., {et~al.} 2006, Annales Geophysicae,
  24, 2619

\bibitem[{{Ofman} \& {Thompson}(2011)}]{ofman2011}
{Ofman}, L., \& {Thompson}, B.~J. 2011, \apjl, 734, L11

\bibitem[{{Otto} \& {Fairfield}(2000)}]{otto2000}
{Otto}, A., \& {Fairfield}, D.~H. 2000, \jgr, 105, 21175

\bibitem[{{Ryu} {et~al.}(2000){Ryu}, {Jones}, \& {Frank}}]{ryu2000}
{Ryu}, D., {Jones}, T.~W., \& {Frank}, A. 2000, \apj, 545, 475

\bibitem[{{Ryutova}(2015)}]{ryuto2015}
{Ryutova}, M. 2015, {Physics of Magnetic Flux Tubes},
  doi:10.1007/978-3-662-45243-1

\bibitem[{{Ryutova} {et~al.}(2010){Ryutova}, {Berger}, {Frank}, {Tarbell}, \&
  {Title}}]{ryutova2010}
{Ryutova}, M., {Berger}, T., {Frank}, Z., {Tarbell}, T., \& {Title}, A. 2010,
  \solphys, 267, 75

\bibitem[{{Wu}(1986)}]{wu1986}
{Wu}, C.~C. 1986, \jgr, 91, 3042

\bibitem[{{Zaliznyak} {et~al.}(2003){Zaliznyak}, {Keppens}, \&
  {Goedbloed}}]{zaliz2003}
{Zaliznyak}, Y., {Keppens}, R., \& {Goedbloed}, J.~P. 2003, Physics of Plasmas,
  10, 4478

\bibitem[{{Zhelyazkov} {et~al.}(2015{\natexlab{a}}){Zhelyazkov}, {Chandra},
  {Srivastava}, \& {Mishonov}}]{zhely2015b}
{Zhelyazkov}, I., {Chandra}, R., {Srivastava}, A.~K., \& {Mishonov}, T.
  2015{\natexlab{a}}, \apss, 356, 231

\bibitem[{{Zhelyazkov} {et~al.}(2015{\natexlab{b}}){Zhelyazkov},
  {Zaqarashvili}, \& {Chandra}}]{zhely2015a}
{Zhelyazkov}, I., {Zaqarashvili}, T.~V., \& {Chandra}, R. 2015{\natexlab{b}},
  \aap, 574, A55

\end{thebibliography}

\end{document}